\documentclass[aps,prl,preprint,superscriptaddress,showpacs,showkeys]{revtex4}
\usepackage{amssymb}
\usepackage{amsmath}
\usepackage{epsfig}
\usepackage{psfrag}

\usepackage{caption2}

\newcommand{\beq}[0]{\begin{equation}}
\newcommand{\eeq}[0]{\end{equation}}

\begin{document}

\title{Risk measures with non-Gaussian fluctuations}

\author{G. Bormetti}
\affiliation{Dipartimento di Fisica Nucleare e Teorica, Universit\`a di Pavia\\
Via A. Bassi 6, 27100, Pavia, Italy}
\affiliation{Istituto Nazionale di Fisica Nucleare, sezione di Pavia\\
Via A. Bassi 6, 27100, Pavia, Italy}
\author{E. Cisana}
\affiliation{Dipartimento di Fisica Nucleare e Teorica, Universit\`a di Pavia\\
Via A. Bassi 6, 27100, Pavia, Italy}
\affiliation{Istituto Nazionale di Fisica Nucleare, sezione di Pavia\\
Via A. Bassi 6, 27100, Pavia, Italy}
\author{G. Montagna}
\affiliation{Dipartimento di Fisica Nucleare e Teorica, Universit\`a di Pavia\\
Via A. Bassi 6, 27100, Pavia, Italy}
\affiliation{Istituto Nazionale di Fisica Nucleare, sezione di Pavia\\
Via A. Bassi 6, 27100, Pavia, Italy}
\affiliation{Istituto Universitario di Studi Superiori (IUSS)\\
Via Luino 4, 27100, Pavia, Italy}
\author{O. Nicrosini}
\affiliation{Istituto Nazionale di Fisica Nucleare, sezione di Pavia\\
Via A. Bassi 6, 27100, Pavia, Italy}
\affiliation{Istituto Universitario di Studi Superiori (IUSS)\\
Via Luino 4, 27100, Pavia, Italy}

\date{\today}

\noindent
\begin{abstract}
  Reliable calculations of financial risk require that the fat-tailed nature 
  of prices changes is included in risk measures. To this end, a non-Gaussian 
  approach to financial risk management is presented, modeling the power-law
  tails of the returns distribution in terms of a Student-$t$ (or Tsallis) distribution. 
  Non-Gaussian closed-form solutions for Value-at-Risk and Expected Shortfall are obtained and
  standard formulae known in the literature under the normality assumption are
  recovered as a special case. The implications of the approach for risk management
  are demonstrated through an empirical analysis of financial time series from
  the Italian stock market. Detailed comparison with the results of the widely used 
  procedures of quantitative finance, such as parametric normal approach, RiskMetrics 
  methodology and historical simulation, as well as with previous findings in the 
  literature, are shown and commented. Particular attention is paid to quantify the size of the errors 
  affecting the risk measures obtained according to different methodologies, by employing a 
  bootstrap technique.
\end{abstract}

\keywords{Econophysics; Financial risk; Risk measures; Fat-tailed distributions; Bootstrap}
\pacs{02.50.Ey - 05.10.Gg - 89.75.-k}

\maketitle

A topic of increasing importance in quantitative finance is the 
development of reliable methods of measuring and controlling 
financial risks. Among the different sources of risk, market risk, which
concerns the hazard of losing money due to the fluctuations
of the prices of those instruments entering a financial portfolio, is particularly relevant.

In the financial industry today, the most widely used measure to manage 
market risk is Value-at-Risk (VaR) \cite{jorion,bouchaud_potters}. In short, VaR refers to the maximum
potential loss over a given period at a certain confidence level and can be used
to measure the risk of individual assets and portfolios of assets as well. Because
of its conceptual simplicity, VaR has become a standard component in the
methodology of academics and financial practitioners. 
However, as discussed in the literature~\cite{bouchaud_potters,acerbietal}, VaR suffers from some inconsistencies: first, it can 
violate the sub-additivity rule for portfolio risk, which is a required property for
any consistent measure of risk, and, secondly, it doesn't quantify the typical loss
incurred when the risk threshold is exceeded. To overcome the drawbacks of
VaR, the Expected Shortfall (or Conditional VaR) is introduced, and sometimes
used in financial risk management, as a more coherent measure of risk.

Three main approaches are known in the literature and used in practice 
for calculating VaR and Expected Shortfall. The first method, called parametric
or analytical, consists in 
assuming some probability distribution function for price changes and 
calculating the risk measures as closed-form solutions. 
Actually, it is well known that empirical price returns, especially in the limit of
high frequency, do not follow the Gaussian paradigm
and are characterized by heavier tails and a higher peak than a normal 
distribution. In order to capture the leptokurtic (fat-tailed) nature of price returns, 
the historical simulation method is often used. 
It employs recent historical data and risk measures are derived from the percentiles
of the distribution of real data.
A third approach consists in Monte Carlo simulations of the stochastic dynamics of a given model
for stock price returns and in calculating risk measures according to Monte Carlo
statistics.

Actually, reliable and possibly fast methods to calculate financial
risk are strongly demanded. Inspired by this motivation, the aim of this paper is
to present a non-Gaussian approach to market risk management and to describe its
potentials, as well as limitations, in comparison with standard procedures used
in financial analysis. To capture the excess of kurtosis of empirical data with respect
to the normal distribution, the statistics of price changes is modeled in terms of a Student-$t$ 
distribution (also known as Tsallis distribution~\cite{gellmann-tsallis}), which is known to 
approximate with good accuracy the distribution derived from market data at a given time 
horizon~\cite{bouchaud_potters,mantegna_stanley}.
We present, in the spirit of a parametric approach, closed-form expressions 
for the risk measures (VaR and ES) and critically investigate the implications of our 
non-Gaussian analytical solutions on the basis of an empirical analysis of financial data. 
Moreover, we perform detailed comparisons with the results of widely used procedures in
finance. Particular attention is paid to quantify the size of the errors affecting 
the various risk measures, by employing a bootstrap technique.


\section{Non-Gaussian risk measures}
\label{s:closed}

Value-at-Risk, usually denoted as $\Lambda^{\star}$, is defined as the maximum potential loss 
over a fixed time horizon $\Delta t$ for a given significance level $\mathcal{P}^{\star}$ 
(typically 1$\%$ or 5$\%$). In terms of price changes $\Delta S$, or, equivalenty, of returns 
$R\doteq\Delta S/S$, VaR can be computed as follows
\beq
  \label{eq:var}
  \mathcal{P}^\star \doteq \int_{-\infty}^{-\Lambda^{\star}} \mathrm{d}\Delta S~\tilde P_{\Delta t}
  (\Delta S)= S \int_{-\infty}^{-\Lambda^{\star}/ S}\mathrm{d}R~P_{\Delta t}(R),
\eeq
where $\tilde P_{\Delta t} (\Delta S)$ and $P_{\Delta t}(R)$ are the probability density functions (pdfs)
for price changes and for returns over a time horizon $\Delta t$, respectively. 
Actually, VaR represents the standard measure used to quantify market risk because 
it aggregates several risk component into a single number. 
In spite of its conceptual simplicity, VaR shows some drawbacks, as mentioned above.

A quantity that does not suffer of these disadvantages is the so called Expected Shortfall (ES) or 
Conditional VaR (CVaR), $E^{\star}$, defined as
\beq
  \label{eq:ES}
  E^{\star} \doteq \frac{1}{\mathcal{P}^\star}\int_{-\infty}^{-\Lambda^\star}\mathrm{d}
  \Delta S~(-\Delta S)~\tilde P_{\Delta t}(\Delta S) = \frac{S}{\mathcal{P}^\star}
  \int_{-\infty}^{-\Lambda^\star/S}\mathrm{d} R~(-R)~P_{\Delta t}(R),
\eeq
with $\mathcal{P}^\star$ and $\Lambda^{\star}$ as in Eq. (\ref{eq:var}).

Assuming returns as normally distributed, i.e. $R\sim\mathcal{N}(m,\sigma^2)$, VaR and ES analytical 
expressions reduce to the following closed-form formulae
\beq
  \label{eq:var_gauss}
  \Lambda^{\star} = -mS_0 + \sigma S_0\sqrt{2}~\mathrm{erfc}^{-1}(2\mathcal{P}^\star)
\eeq
and
\beq
  \label{eq:ES_gauss}
  E^{\star} = -m S_0 + \frac{\sigma S_0}{\mathcal{P}^\star}\frac{1}
  {\sqrt{2\pi}}\exp\{-[\mathrm{erfc}^{-1}(2\mathcal{P}^\star)]^2\},
\eeq
where $\mathrm{erfc}^{-1}$ is the inverse of the complementary error function~\cite{nr}. Note that 
expressions~(\ref{eq:var_gauss}) and (\ref{eq:ES_gauss}) are linear with respect to the spot price $S_0$.

However, it is well known in the literature~\cite{bouchaud_potters,mantegna_stanley,mandelbrot} that 
the normality hypothesis is often inadequate for daily returns due to the leptokurtic nature of 
empirical data. For this reason, a better agreement with data is obtained using fat-tailed distributions, 
such as truncated L\'evy distributions~\cite{mantegna_stanley,mantegna_prl} or Student-$t$ ones. 
In order to characterize the excess of kurtosis, we model the returns using a Student-$t$ distribution defined as 
\beq
  \label{eq:student}
  \mathcal{S}^{\nu}_{m,a} (R) = \frac {1}{B(\nu/2,1/2)} \frac {a^{\nu}} {[a^{2}+(R-m)^{2}]^{\frac{\nu+1}{2}}},
\eeq
where $\nu\in (1,+\infty)$ is the tail index and $B(\nu/2,1/2)$ is the beta function. 
It is easy to verify that, for $\nu > 2$, the variance 
is given by $\sigma^{2} = a^2/(\nu - 2)$, while, for $\nu > 4$, 
the excess kurtosis reduces to $k = 6/(\nu - 4)$. Under this assumption, we obtain closed-form generalized expression for VaR and ES given by
\beq
  \label{eq:var_stud}
  \Lambda^\star = -mS_0+\sigma S_0\sqrt{\nu-2}~\sqrt{\frac{1-\lambda^\star}{\lambda^\star}}
\eeq
and
\beq
  \label{eq:ES_stud}
  E^\star = -m S_0 + \frac{\sigma S_0}{\mathcal{P}^\star B(\nu/2,1/2)}
  \frac{\sqrt{\nu-2}}{\nu-1}~[\lambda^\star]^\frac{\nu-1}{2},
\eeq
where $\lambda^\star\doteq I^{-1}_{[\nu/2,1/2]}(2\mathcal{P}^\star)$ and $I^{-1}_{[\nu/2,1/2]}$ 
\begin{figure}[t!]
  \caption{\label{fig:convergence} Convergence of VaR (left) and ES (right) Student-$t$ formulae 
    toward Gaussian.}
      \psfrag{1p}[l]{\tiny 1\%}
      \psfrag{5p}[l]{\tiny 5\%}
      \psfrag{n}[rb]{\tiny Normal~~}
      \psfrag{2}[rb]{\tiny $\nu = 2.75$}
      \psfrag{3}[rb]{\tiny $\nu = 3.50$}
      \psfrag{4}[rb]{\tiny $\nu = 4.50$}
      \psfrag{1}[rb]{\tiny $\nu = 100~$}
      \psfrag{Lstar}{\footnotesize $\Lambda^\star$}
      \psfrag{Pstar}[t]{\footnotesize $\mathcal{P}^\star$} 
      \includegraphics[scale = 0.55]{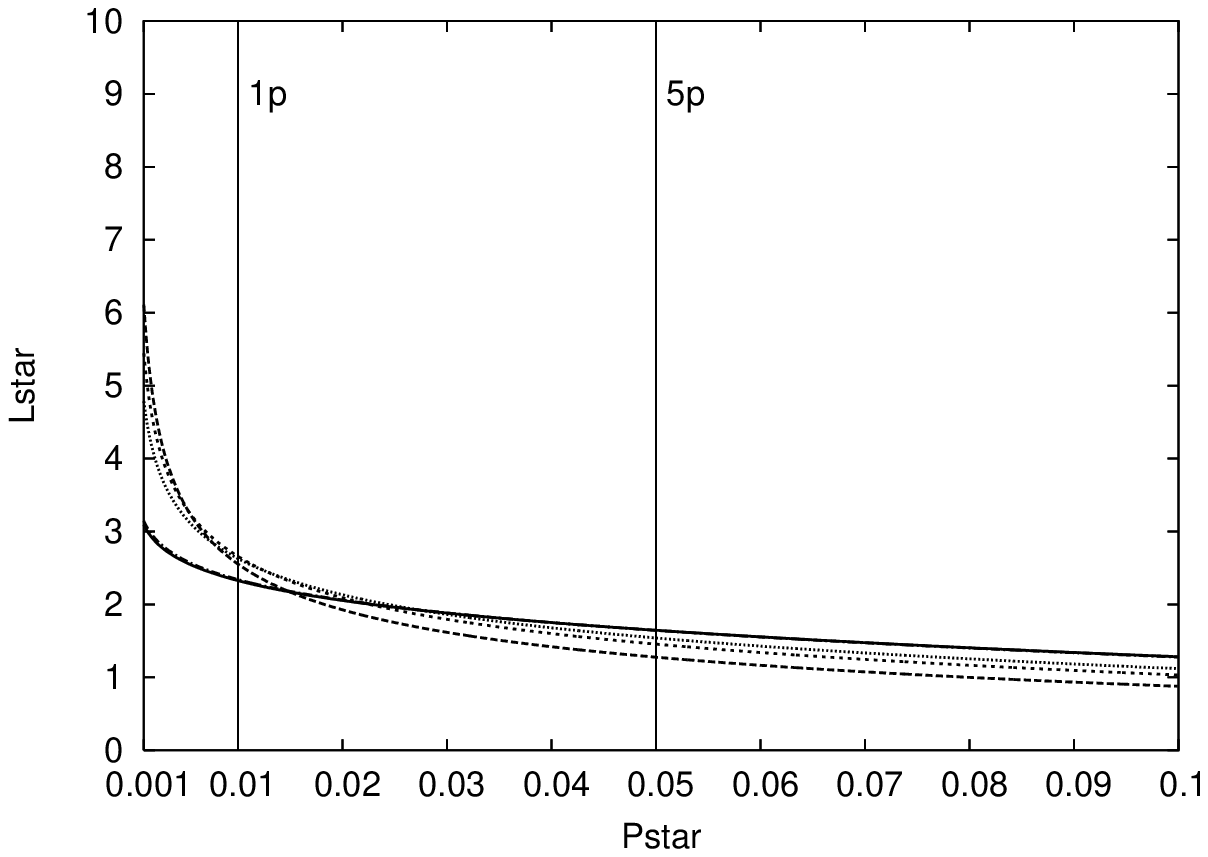}
      \psfrag{1p}[l]{\tiny 1\%}
      \psfrag{5p}[l]{\tiny 5\%}
      \psfrag{n}[rb]{\tiny Normal~~}
      \psfrag{2}[rb]{\tiny $\nu = 2.75$}
      \psfrag{3}[rb]{\tiny $\nu = 3.50$}
      \psfrag{4}[rb]{\tiny $\nu = 4.50$}
      \psfrag{1}[rb]{\tiny $\nu = 100~$}
      \psfrag{Estar}{\footnotesize $\mathrm{E}^\star$}
      \psfrag{Pstar}[t]{\footnotesize $\mathcal{P}^\star$}
      \includegraphics[scale = 0.55]{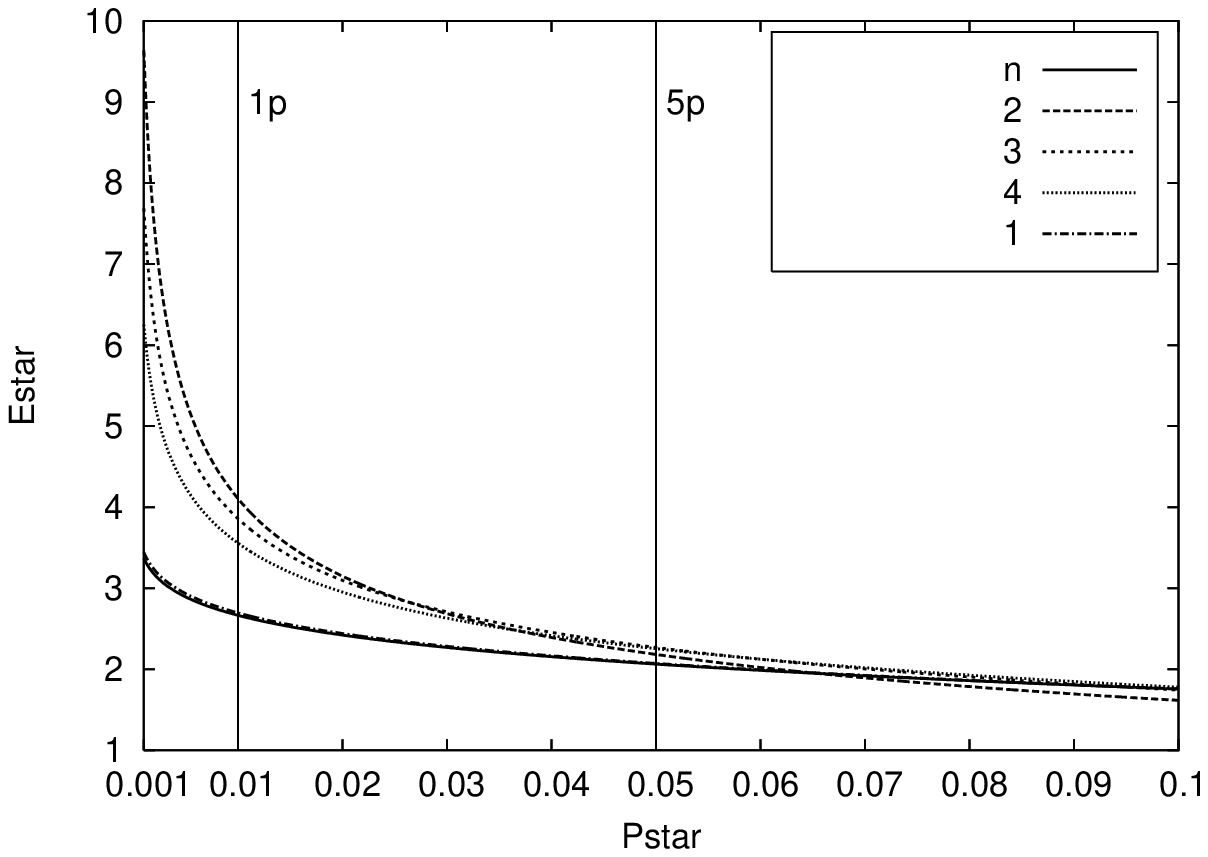}
 \end{figure}
is the inverse of the incomplete beta function~\cite{nr}.

\noindent As shown in Fig.~\ref{fig:convergence}, we have checked numerically the convergence of 
formulae~(\ref{eq:var_stud}) and (\ref{eq:ES_stud}) to the Gaussian results (\ref{eq:var_gauss}) and 
(\ref{eq:ES_gauss}) using different values of tail index $\nu$ (2.75,3.5,4.5,100). As expected, the 
points corresponding to $\nu=100$ are almost coincident with the Gaussian predictions, demonstrating that 
our results correctly recover the Gaussian formulae as a special case. 

We observe that each line, corresponding to a fixed $\nu$, crosses over the Gaussian one
for a certain $\mathcal{P}^\star$. 
In the light of this observation, we report  in Table~\ref{tab:cross} the values of $\nu_{\mathrm{cross}}$ 
corresponding to a given $\mathcal{P}^\star$ for both VaR and ES. As can be observed, the growth of 
$\nu_{\mathrm{cross}}$ with $\mathcal{P}^\star$ is very rapid for VaR, while for ES  and for usually adopted 
significance values, $\nu_{\mathrm{cross}}$ keeps in the interval $[2.09,2.51]$. From 
this point of view, VaR and ES are quite different measures of risk, 
since the crossover values for the latter are much more stable than those associated to the first one. 
This result can be interpreted as a  consequence of ES as a more coherent risk measure than VaR.
\begin{table}[h!]
  \caption{\label{tab:cross} Values of $\nu$ crossover for VaR and ES for different $\mathcal{P}^\star$. }
  \begin{center}
    \begin{tabular}{@{}lccccc@{}}
      \hline
      $\mathcal{P}^\star$ & 1\% & 2\% & 3\% & 4\% & 5\% \\
      \hline 
      \hline
      $\nu_{\mathrm{cross}}$(VaR) & 2.44 & 3.21 & 5.28 & 32.38 & $\gg 100$  \\
      $\nu_{\mathrm{cross}}$(ES) & 2.09 & 2.18 & 2.28 & 2.38 & 2.51  \\
      \hline
    \end{tabular}
  \end{center}
\end{table}
%


\section{Empirical analysis} 
\label{s:data}

The data sets used in our analysis consist of two financial time series, composed of $N=1000$ daily returns,
from the Italian stock market: one is a collection of data from the Italian asset Autostrade SpA (from May 15$^{th}$ 
2001 to May 5$^{th}$ 2005), while the other one corresponds to the financial index Mib30 (from March 27$^{th}$ 2002 to 
March 13$^{th}$ 2006). The data have been freely downloaded from Yahoo Finance \cite{yahoo}. Other examples of
analysis of Italian stock market data can be found in \cite{bormetti}.

In order to balance between accuracy and computational time, we estimate mean $m$ and variance
$\sigma$ from empirical moments. With this standard procedure, we measure daily volatilities of the order of $1\%$
for both time series: $\sigma_{Autostrade} = 1.38\%$ and $\sigma_{Mib30} = 1.16\%$. Moreover, we find quite 
negligible means: $m_{Autostrade} = 0.12\%$ and $m_{Mib30} = 0.02\%$. Using the above $m$ and $\sigma$ values, we derive 
a standardized vector (with zero mean and unit variance) $\mathbf{r}\doteq(r_{t},\ldots,r_{t-N+1})$, where $r_{t-i}\doteq (R_{t-i}-m)/\sigma$ for $i = 0,\ldots,N-1$. In order to find the best value for the tail parameter $\nu$, 
we look for the argument that minimizes the negative log-likelihood, according to the formula 
\beq\label{eq:neglog} 
\nu = \mathrm{argmin}_{~\nu>2} \left[-\sum_{i=0}^{N-1}\log \mathcal{S}^{\nu}_{0,\sqrt{\nu-2}}(r_{t-i})\right],
\eeq
where the constraint $\nu>2$ prevents the variance to be divergent and $\mathcal{S}^{\nu}_{0,\sqrt{\nu-2}}$ is as in 
Eq.~(\ref{eq:student}), with $m=0$ and $a=\sqrt{\nu-2}$. We remark that the beta function $B(\nu/2,1/2)$ only 
admits an integral representation and therefore we implemented a numerical algorithm to search for the minimum 
and solve the optimization problem. We measure tail parameter of $2.91$ for Autostrade SpA and $3.22$ for Mib30. 
These values confirm the strong leptokurtic nature of the returns distributions, both for single asset and market index.  

For completeness, it is worth mentioning that other approaches are discussed in the literature to model with 
accuracy the tail exponent of the returns cdfs and are based on Extreme Value Theory \cite{frey} and Hill's estimator 
\cite{epjb,clementi}.


\section{Comparison of different risk methodologies}
\label{s:risk}

In this Section we present a comparison of the results obtained estimating the market 
risk through VaR and ES according to different methodologies. The standard approach is based on the normality assumption for the distribution of the returns.
As discussed above, we limit our analysis to $1000$ daily data and we 
estimate the volatility using the empirical second moment (the effect of the mean is negligible). 
In order to avoid the problem of a uniform weight for the returns, RiskMetrics 
introduces the use of an exponential weighted moving average of 
squared returns according to the formula~\cite{mina_xiao}
\beq\label{eq:riskm}
  \sigma_{t+1\mid t}^2 \doteq \frac{1-\lambda}{1-\lambda^{N+1}}\sum_{i=0}^{N-1} \lambda^i (R_{t-i}-m)^2,
\eeq
where $\lambda \in (0,1]$ is a decay factor. 
The choice of $\lambda$ depends on the time horizon and, for $\Delta t = 1$ day, 
$\lambda = 0.94$ is the usually adopted value \cite{mina_xiao}. $\sigma_{t+1\mid t}$ represents volatility estimate at time $t$ conditional on 
the realized $\mathbf{R}$. %
In order to relax standard assumption about the return pdf without loosing the advantages coming from
a closed-form expression, we presented above generalized 
formulae for VaR and ES based on a Student-$t$ modeling of price returns. 
As a benchmark of all our results, we also quote VaR and ES estimates following a historical approach, 
which is a procedure widely used in the practice. According to this
approach, after ordering the $N$ data in increasing order, we consider the $[N\mathcal{P^\star}]^\mathrm{th}$
return $R_{([N\mathcal{P^\star}])}$ as an estimate for VaR 
and the empirical mean over first $[N\mathcal{P^\star}]$ returns as an estimate for ES.

\begin{figure}[t!]
  \caption{\label{fig:boot} Bootstrap histograms for tail index $\nu$ (left) and for RiskMetrics volatility
    proxy $\sigma_{t+1|t}$ (right) for Autostrade SpA ($M=10^{3}$ bootstrap copies)}
  \psfrag{theta}[c]{\footnotesize $\theta = \nu$}
  \psfrag{nu}[lb]{\tiny~$\nu$}
  \psfrag{nu^*}[c]{\footnotesize $\nu^*$}
  \psfrag{Frequency}[c]{\footnotesize$\mathrm{Frequency}$}
  \psfrag{a}[lb]{\tiny~$\nu^{*}_{16\%}$}
  \psfrag{b}[lb]{\tiny~$\nu^{*}_{b}$}
  \psfrag{ca}[lb]{\tiny~$\nu^*_{84\%}$}
  \includegraphics[scale = 0.55]{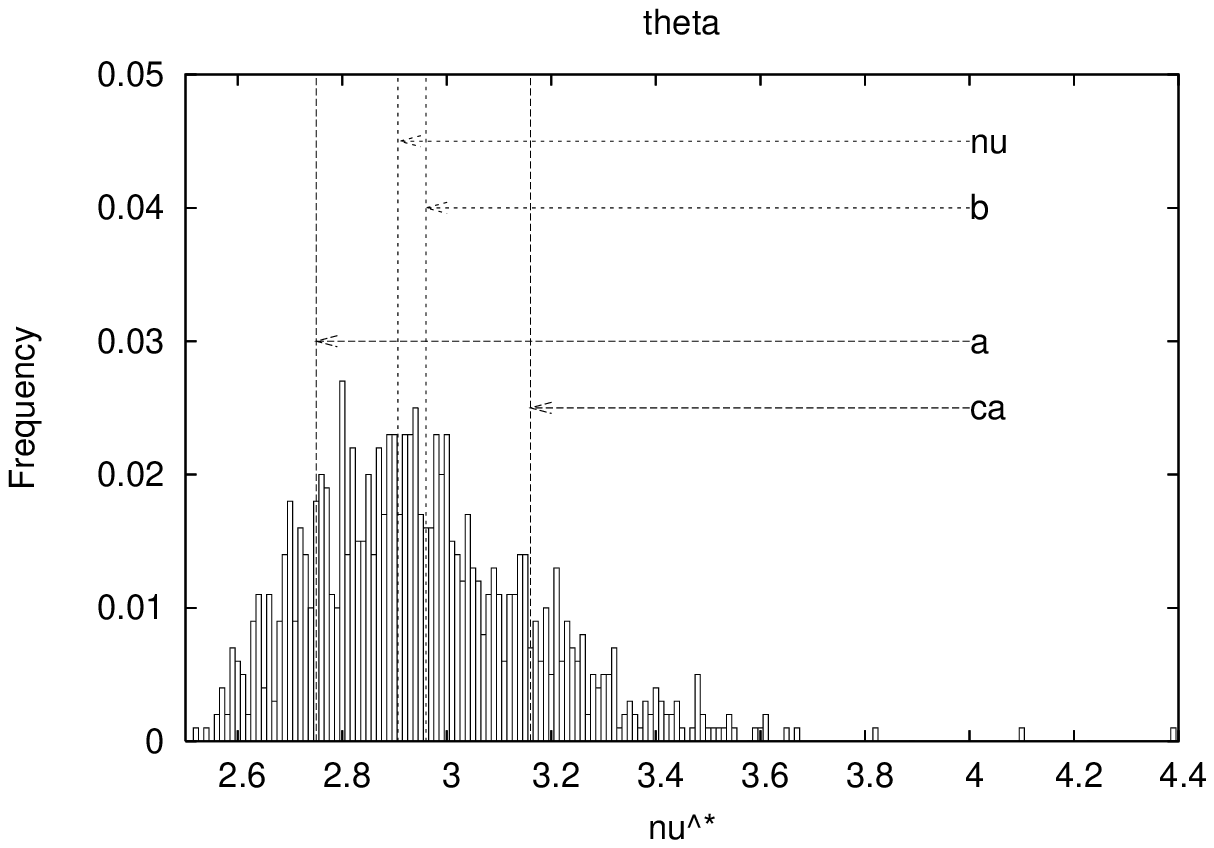}
  \psfrag{theta}[c]{\footnotesize$\theta = \sigma_{t+1|t}$}
  \psfrag{vre}[lb]{\tiny~$\sigma_{t+1|t}$}
  \psfrag{vre^*}[c]{\footnotesize$\sigma_{t+1|t}^{*}$}
  \psfrag{Frequency}[c]{\footnotesize$\mathrm{Frequency}$}
  \psfrag{a}[lb]{\tiny~$\sigma_{(t+1|t)16\%}^{*}$}
  \psfrag{b}[lb]{\tiny~$\sigma_{t+1|t}^*$}
  \psfrag{ca}[lb]{\tiny~$\sigma_{(t+1|t)84\%}^{*}$}
  \includegraphics[scale = 0.55]{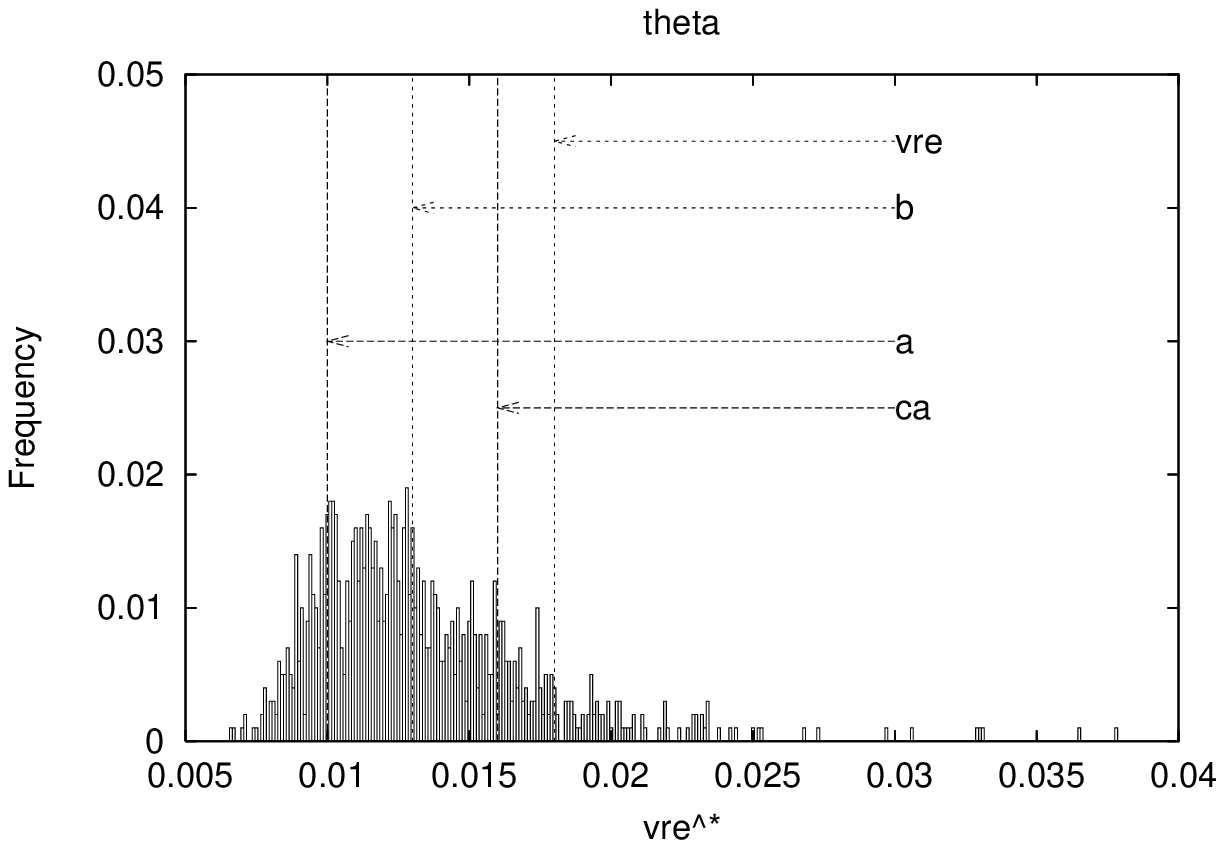}
\end{figure}

\begin{figure}[t!]
  \caption{\label{fig:es_var}
    VaR $\Lambda^\star$ (upper panel) and ES $\mathrm{E}^\star$ (lower panel) central values with 68\% CL intervals for Autostrade SpA (left) and 
    for Mib30 (right).}
      \psfrag{lstar}[c]{\footnotesize $\Lambda^\star$}
      \psfrag{ML}[c]{\tiny $\mathrm{Student-}t$} 
      \psfrag{N}[c]{\tiny $\mathrm{Normal}$} 
      \psfrag{H}[c]{\tiny $\mathrm{Historical}$}
      \psfrag{EW}[c]{\tiny $\mathrm{~~~~~RiskMetrics}$}
      \includegraphics[scale = 0.55]{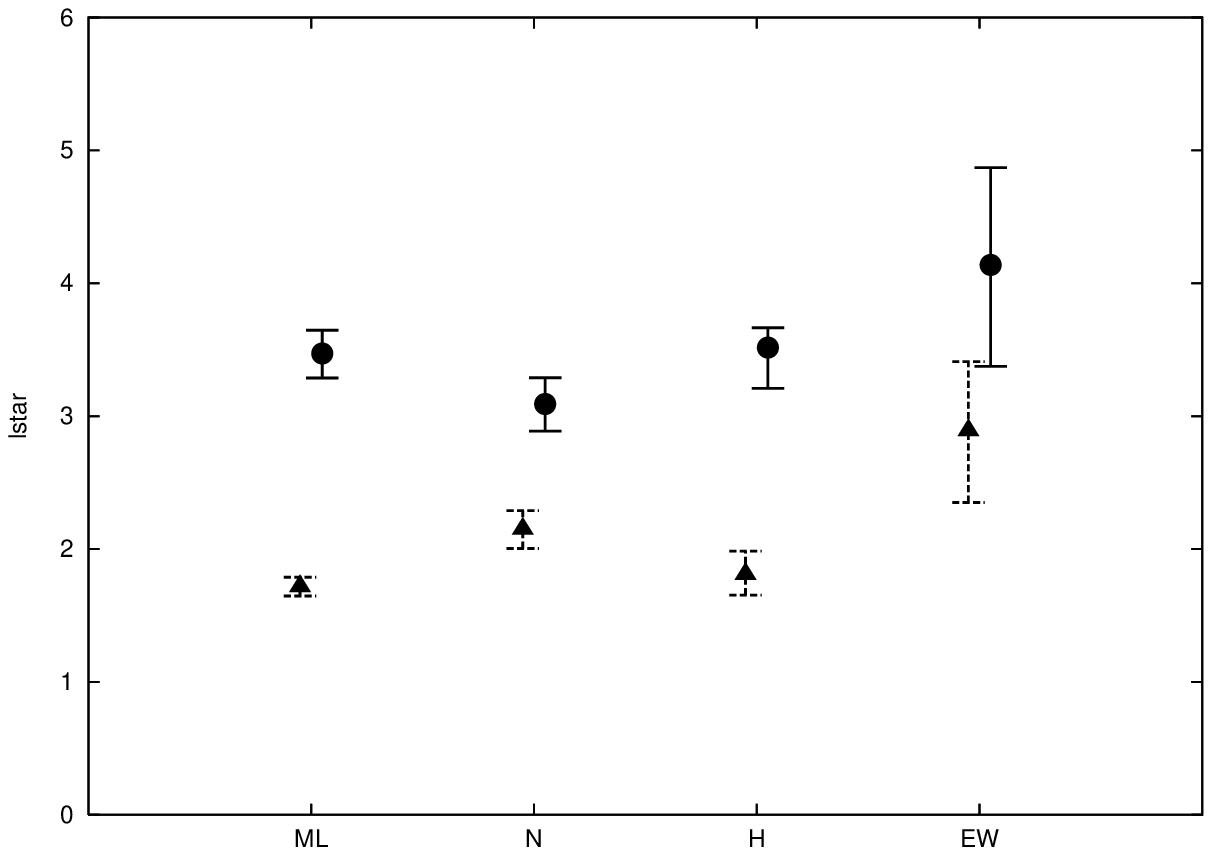}
      \psfrag{lstar}[c]{\footnotesize $~$}
      \psfrag{ML}[c]{\tiny $\mathrm{Student-}t$} 
      \psfrag{N}[c]{\tiny $\mathrm{Normal}$} 
      \psfrag{H}[c]{\tiny $\mathrm{Historical}$}
      \psfrag{EW}[c]{\tiny $\mathrm{~~~~~RiskMetrics}$} 
      \psfrag{1}[rb]{\tiny $\mathcal{P}^\star 1\%$}
      \psfrag{5}[rb]{\tiny $\mathcal{P}^\star 5\%$}
      \includegraphics[scale = 0.55]{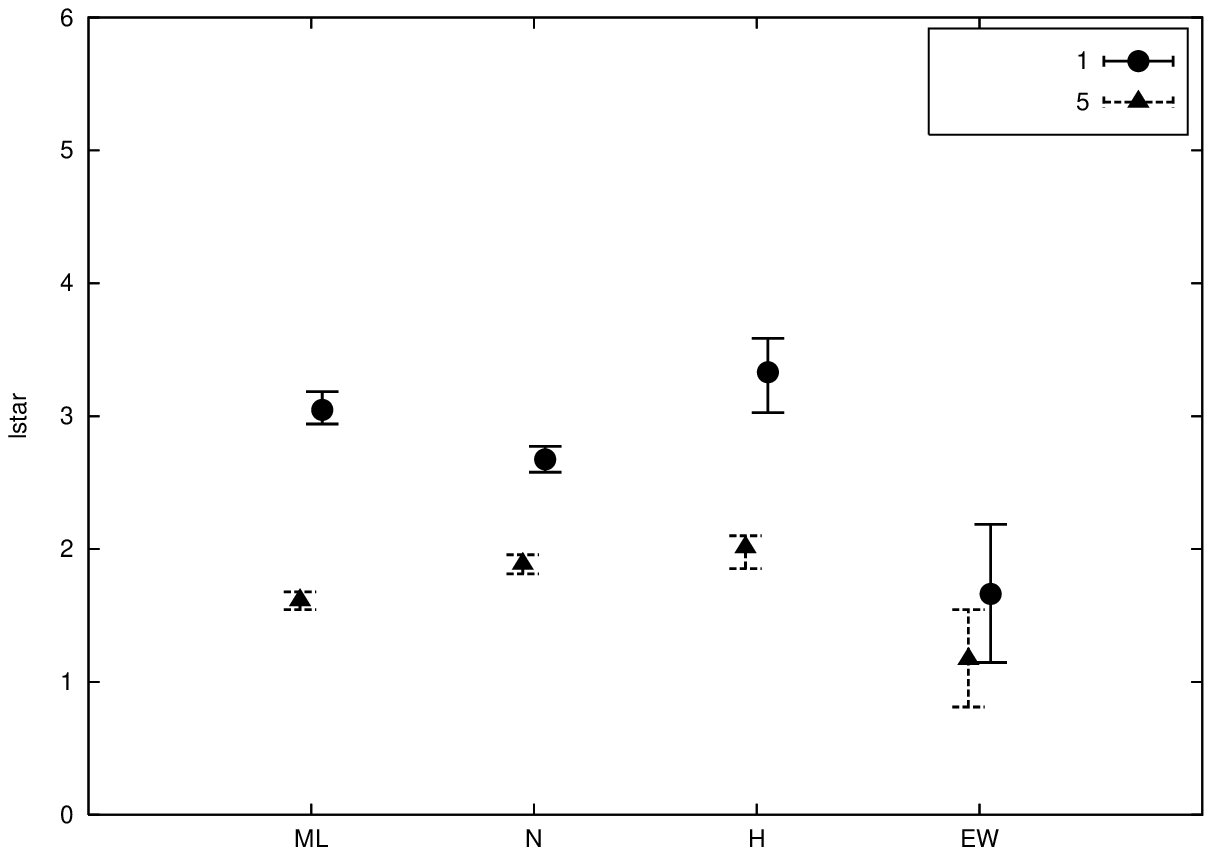}
      \psfrag{estar}[c]{\footnotesize $\mathrm{E}^\star$}
      \psfrag{ML}[c]{\tiny $\mathrm{Student-}t$} 
      \psfrag{N}[c]{\tiny $\mathrm{Normal}$} 
      \psfrag{H}[c]{\tiny $\mathrm{Historical}$}
      \psfrag{EW}[c]{\tiny $\mathrm{~~~~~RiskMetrics}$}
      \psfrag{AUTS}[ct]{\footnotesize $\mathrm{Autostrade}$}
      \includegraphics[scale = 0.55]{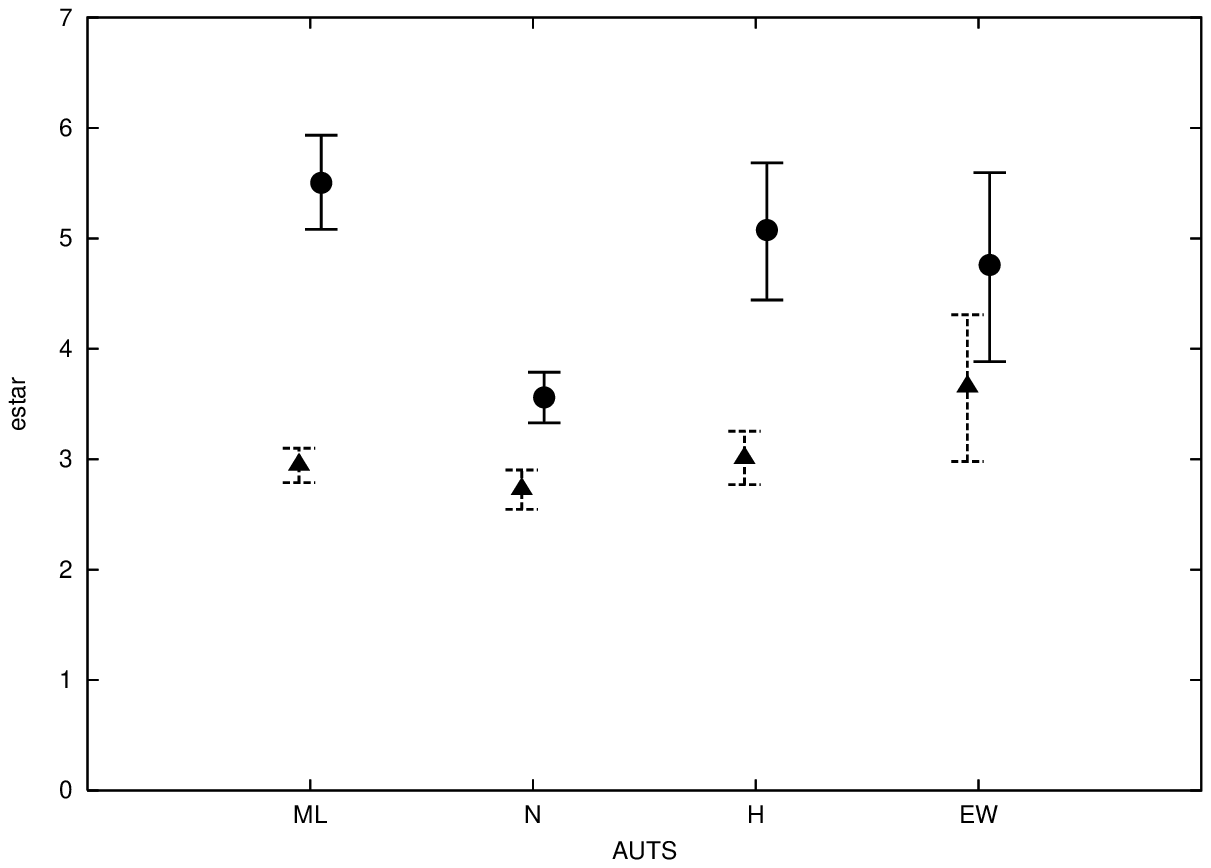}
      \psfrag{estar}[c]{\footnotesize $~$}
      \psfrag{ML}[c]{\tiny $\mathrm{Student-}t$} 
      \psfrag{N}[c]{\tiny $\mathrm{Normal}$} 
      \psfrag{H}[c]{\tiny $\mathrm{Historical}$}
      \psfrag{EW}[c]{\tiny $\mathrm{~~~~~RiskMetrics}$}
      \psfrag{MIB3}[ct]{\footnotesize $\mathrm{Mib30}$}
      \includegraphics[scale = 0.55]{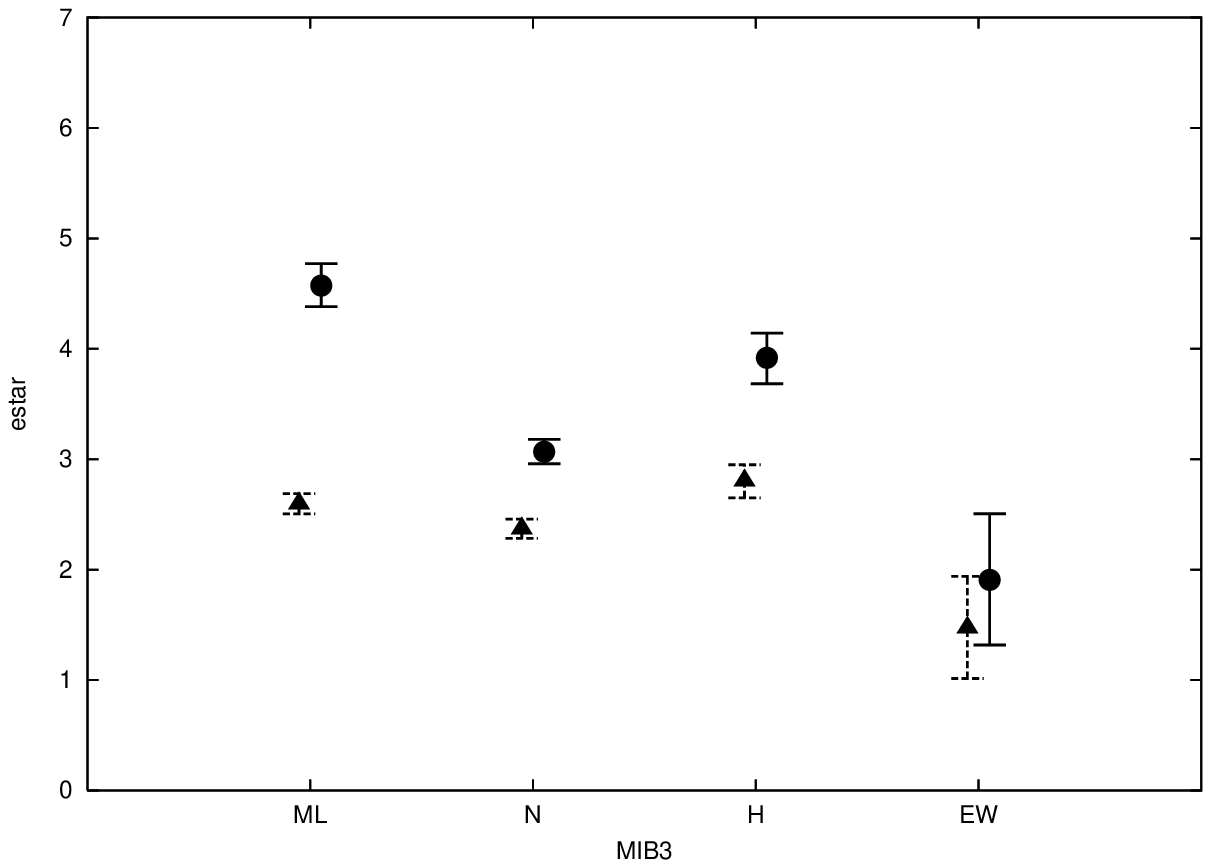}
  \end{figure}

At a variance with respect to previous investigations \cite{mattedi_etal,pafka_kondor}, 
we also provide 68\% confidence level (CL) intervals associated to the parameters, 
to estimate VaR and ES dispersion.
To this extent, we implement a bootstrap technique \cite{efron_tibshirani}. 
Given the $N$ measured returns, we generate $M=1000$ synthetic copies of $\mathbf{R}$,
$\{\mathbf{R}^*_j\}$, with $j=1,\ldots,M$, by random sampling with replacement according to the probability 
$p=(1/N,\ldots,1/N)$. For each $\mathbf{R}^*_j$ we estimate the quantities of interest $\theta$ and we obtain 
bootstrap central values as follow 
\beq\label{eq:mean_boot}
  \theta^*_b\doteq\frac{1}{M}\sum_{j=1}^M \theta^*_j.
\eeq
We define the $1-2\alpha$ CL interval as $[\theta_\alpha^*,\theta_{1-\alpha}^*]$, with $\theta_a^*$ such that 
$P(\theta^*\leq \theta^*_a)=a$ and $a=\alpha,1-\alpha$. 68\% CL implies $\alpha=16$\%. In 
Fig.~\ref{fig:es_var}, Tables~\ref{tab:boot} and \ref{tab:var}
we quote results according to $\theta-(\theta_b^*-\theta_\alpha^*)+(\theta_{1-\alpha}^*-\theta_b^*)$. 
In this way, we use the bootstrap approach in order to estimate the dispersion of the quantity of 
interest around the measured value $\theta$. In Fig.~\ref{fig:boot} we present, as example, the
histogram of bootstrap values of tail index $\nu$ for Autostrade SpA. We observe that the empirical 
value $\nu = 2.91$ is close to the bootstrap central value $\nu^{*}_{b} = 2.96$.
We also include $68\%$ CL interval ($\nu^{*}_{16\%}=2.75$ and $\nu^*_{84\%}=3.16$), in order to quantify the dispersion 
around $\nu^{*}_{b}$. 
\begin{table}[!]
  \caption{\label{tab:boot}Parameters values and bootstrap estimates for the 68\% CL intervals.
     }
  \begin{center}
    \begin{tabular}{@{}lccccc@{}}
      \hline
      & $m$ & $\sigma$ & $\sigma_{t+1\mid t}$ & $\nu$ & $R_{(10)}$ \\
      \hline 
      \hline
      Autostrade & $0.12^{+0.04}_{-0.05}\%$  & $1.38^{+0.08}_{-0.10}\%$  & $1.83^{+0.31}_{-0.33}\% $ & $2.91^{+0.20}_{-0.21}$ & $-3.51_{-0.15}^{+0.31}\%$\\ 
          Mib30      & $0.02^{+0.03}_{-0.04}\%$  & $1.16^{+0.03}_{-0.05}\%$  & $0.72^{+0.22}_{-0.22}\%$ & $3.22^{+0.15}_{-0.16}$ & $-3.33^{+0.30}_{-0.25}\%$ \\  
      \hline
    \end{tabular}
  \end{center}
\end{table}

As a rule of thumb, we consider the bootstrap approach accurate when, given a generic parameter, the difference between 
its empirical value and the bootstrap central value estimate is close to zero and 68\% CL interval is almost symmetric. 
In our numerical investigation, we found a systematic non zero bias for $\theta = \sigma_{t+1\mid t}$. In Fig.~\ref{fig:boot},
$\sigma_{t+1\mid t} = 1.83\%$ while $\sigma_{t+1\mid t}^{*} = 1.32\%$, so we measured a bias of order 0.005. It is 
worth noticing the positive skewness of the histogram. From Table~\ref{tab:boot} 
it is quite evident the asymmetry of $R_{([N\mathcal{P^\star}])}$.
Therefore, we can consider the corresponding CL intervals as a first approximation of the right ones~\cite{efron_tibshirani}.

In Fig.~\ref{fig:es_var} we show VaR and ES central values and 68\% CL bars 
for Autostrade SpA and Mib30, corresponding to 1\% and 5\% significance level and according 
to the four methodologies previously described. In Tables~\ref{tab:var} we detail all the numerical results.
 As already noted in Ref.~\cite{pafka_kondor}, at 5\% significance level
Student-$t$ and Normal approaches are substantially equivalent, but here such a statement
sounds more statistically robust, thanks to the bootstrap 68\% confidence levels and to the 
comparison with the historical simulation. 
We note also, from Fig.~\ref{fig:es_var} and Table~\ref{tab:var}, that $\Lambda^\star$ and $\mathrm{E}^\star$ 
central values calculated according to RiskMetrics   
methodology are quite fluctuating and characterized by the largest CL bars.
The decreasing of $\mathcal{P}^\star$ traduces in a major differentiation of the different approaches. 
In general, we obtain the best agreement between 
the Student-$t$ approach and the historical simulation, both for $\Lambda^\star$ and 
$\mathrm{E}^\star$, whereas, as before, the RiskMetrics methodology overestimates or
underestimates the results of the historical evaluation and is affected by rather large 
uncertainties. 

\begin{table}
  \begin{center}
    \caption{\label{tab:var}Estimated VaR and ES values (mean and 68\% CL interval) for 
    1$\%$ and 5$\%$ significance levels 
      from Autostrade SpA and Mib30.} 
  \end{center}
  \begin{center}
    \begin{tabular}{ll|l|l|l|l}
      \hline
      & & Student-$t$ & Normal & Historical & RiskMetrics\\
      \hline
      \hline
      Autostrade & VaR 1\% & $3.472_{-0.185}^{+0.175}$ & $3.091_{-0.204}^{+0.197}$ & $3.516_{-0.306}^{+0.149}$ & $4.138_{-0.764}^{+0.733}$\\
      & VaR 5$\%$ &  $1.717_{-0.071}^{+0.071}$ & $2.150_{-0.145}^{+0.139}$ & $1.810_{-0.156}^{+0.175}$ & $2.890_{-0.540}^{+0.520}$\\
      \hline  
      & ES 1\% & $5.503_{-0.421}^{+0.431}$ & $3.559_{-0.231}^{+0.229}$ & $5.076_{-0.634}^{+0.607}$ & $4.759_{-0.876}^{+0.837}$\\
      & ES 5\% & $2.946_{-0.159}^{+0.153}$ & $2.727_{-0.182}^{+0.175}$ & $3.006_{-0.235}^{+0.248}$ & $3.655_{-0.677}^{+0.653}$\\
      \hline
      Mib30 & VaR 1\% & $3.047_{-0.105}^{+0.106}$ & $2.675_{-0.096}^{+0.097}$ & $3.331_{-0.304}^{+0.255}$ & $1.662_{-0.516}^{+0.524}$\\
      & VaR 5$\%$ & $1.612_{-0.067}^{+0.066}$ & $1.885_{-0.072}^{+0.073}$ & $2.010_{-0.157}^{+0.090}$ & $1.169_{-0.358}^{+0.375}$\\
      \hline 
      & ES 1\% & $4.572_{-0.191}^{+0.199}$ & $3.068_{-0.109}^{+0.111}$ & $3.918_{-0.234}^{+0.223}$ & $1.908_{-0.590}^{+0.599}$\\
      & ES 5\% & $2.596_{-0.091}^{+0.093}$ & $2.369_{-0.086}^{+0.088}$ & $2.804_{-0.155}^{+0.145}$ & $1.471_{-0.458}^{+0.467}$\\
      \hline
    \end{tabular}
  \end{center}
\end{table}


\section{Conclusions}
\label{s:conclusion}

In this paper we have presented a careful analysis of financial market risk measures 
in terms of a non-Gaussian (Student-like) model for price fluctuations.
With the exception of Gaussian ones, the derived closed-form parametric formulae are able 
to capture accurately the fat-tailed nature of financial data and are in good agreement with 
a full historical evaluation.
We also proposed a bootstrap-based technique to estimate the
size of the errors affecting the risk measures derived through the different procedures, 
in order to give a sound statistical meaning to our comparative analysis. 

Possible perspectives concern the extension of our analysis to other time series, different
financial instruments and underlying distributions.

\end{document}